\documentclass[apj]{emulateapj}
\usepackage{amssymb}
\usepackage{epsfig}
\usepackage{longtable} 
\usepackage{amsmath}
\usepackage{amsfonts}
\usepackage{ctable}
\usepackage{graphics}
\usepackage{subfigure}
\usepackage{verbatim}
\begin{document}


\title{Cool Core Bias in Sunyaev-Zel'dovich Galaxy Cluster Surveys}

\author{Henry W. Lin\altaffilmark{1}, Michael McDonald\altaffilmark{2}, Bradford Benson\altaffilmark{3,4,5}, and Eric Miller\altaffilmark{2}}
\email{henrylin@college.harvard.edu}
\altaffiltext{1}{Harvard University, Cambridge, MA 02138, USA}
\altaffiltext{2}{Kavli Institute for Astrophysics and Space Research, MIT, Cambridge, MA 02139, USA}
\altaffiltext{3}{Fermi National Accelerator Laboratory, Batavia, IL 60510-0500}
\altaffiltext{4}{Kavli Institute for Cosmological Physics, University of Chicago, 5640 South Ellis Avenue, Chicago, IL 60637}
\altaffiltext{5}{Department of Astronomy and Astrophysics, University of Chicago, 5640 South Ellis Avenue, Chicago, IL 60637}

\begin{abstract}
Sunyaev-Zel’dovich (SZ) surveys find massive clusters of galaxies by measuring the inverse Compton scattering of cosmic microwave background off of intra-cluster gas. 
The cluster selection function from such surveys is expected to be nearly independent of redshift and cluster astrophysics. 
In this work, we estimate the effect on the observed SZ signal of centrally-peaked gas density profiles (cool cores) and radio emission from the brightest cluster galaxy (BCG) by creating mock observations of a sample of clusters that span the observed range of classical cooling rates and radio luminosities. 
For each cluster, we make simulated SZ observations by the South Pole Telescope and characterize the cluster selection function, but note that our results are broadly applicable to other SZ surveys.
We find that the inclusion of a cool core can cause a change in the measured SPT significance of a cluster between $0.01\%$--$10\%$ at $z > 0.3$, increasing with cuspiness of the cool core and angular size on the sky of the cluster (i.e., decreasing redshift, increasing mass).
We provide quantitative estimates of the bias in the SZ signal as a function of a gas density cuspiness parameter, redshift, mass, and the 1.4 GHz radio luminosity of the central AGN. 
Based on this work, we estimate that, for the Phoenix cluster (one of the strongest cool cores known), the presence of a cool core is biasing the SZ significance high by $\sim$6\%.
The ubiquity of radio galaxies at the centers of cool core clusters will offset the cool core bias to varying degrees. 
\end{abstract}

\keywords{cosmology: observations -- galaxies: cooling flows -- galaxies: clusters}
\maketitle

\section{Introduction} 
	{Galaxy clusters are potentially powerful tools to study dark energy and cosmology} \citep[see review by][]{cosmorev}; whereas standard candles test cosmology on homogeneous scales, clusters trace the growth of inhomogeneity via $N(M,z)$, the cluster number density at a given mass and redshift. 
Accurate cluster cosmology requires not only reliable estimators of clusters properties such as total mass (i.e., $M_{500}$), but also an accurate estimation of the survey selection function.

	A new technique of assembling a nearly unbiased sample of galaxy clusters uses the Sunyaev-Zel'dovich (SZ) effect \citep{szoriginal}: the inverse Compton scattering of photons as they pass through the hot intracluster medium, leading to distorted-spectrum patches in the cosmic microwave background (CMB). One attractive characteristic of SZ surveys is that the SZ signal is virtually redshift independent, as opposed to the $\sim 1/d_{L}^2$ flux dimming in infrared, optical, and X-ray surveys. Recently, large samples ($\gtrsim$100) of galaxy clusters have been assembled using SZ selection that include clusters out to $z \sim 1.5$ \citep[e.g., ][]{planckearly, act, vander10, spt13} including many of the most massive known galaxy clusters \citep[e.g.,][]{mena, foley, nature}.

	{Many potential systematics have been considered in SZ-selected cluster surveys. 
	Simulations have shown that SZ surveys are expected to be relatively insensitive to the effects of non-gravitational physics \citep{nagai06}, projection effects \citep{shawhb}, and the dynamical state of the cluster \citep{mergerbias}.}
	The presence of a ``cool core" --- a central region of over-dense gas accompanied by a drop in temperature --- could also conceivably bias mass estimates. Finally, contamination from radio-loud active galactic nuclei (AGN), generally located in the brightest cluster galaxy, could introduce bias by ``filling in" the SZ decrement \citep{sayers}. 

	Cool core bias is also of particular astrophysical interest, as estimates of the cool core fraction as a function of redshift might lead to valuable insights into the cooling flow problem \citep[see review by][]{fabian94}. Recent studies have found evidence for a lack of dense, cool cores in the centers of high-redshift galaxy clusters \citep[e.g., ][]{V2007,santos,mcdonald11,new}. These results may be evidence for an evolution in the cooling/feedback balance, or may be a result of biases in X-ray and optical surveys \citep[e.g., ][]{sptz}.  Thus, in order to fully understand the evolution of cooling and heating processes in the cores of galaxy clusters, we must understand how our sample selection affects the observed cool core fraction.

	Throughout this paper, we take the fractional matter density $\Omega_m = 0.27$ and the fractional vacuum density $\Omega_\Lambda = 0.73$, and neglect radiation $\Omega_r$ and curvature $\Omega_k$. 
	We introduce two dimensionless forms of the Hubble parameter, $h \equiv H_0 /(100 \, \mathrm{km}\, \mathrm{s}^{-1} \mathrm{Mpc}^{-1}) = 0.7$ and $E(z) \equiv H(z)/H_0$. We define the cluster radius $r_{500}$ such that the average matter density interior to $r_{500}$ is 500 times the critical density $ \rho_\text{cr}$ of the universe as well as the cluster mass $M_{500} = \frac{4}{3} \pi r_{500}^3 \times  500 \rho_\text{cr}$.

\section{Methods}
	In order to measure the influence of cool cores and radio-loud AGN on the SZ signal, we generate one-dimensional (spherically-symmetric) mock galaxy clusters with properties drawn from existing samples of high-mass, relaxed clusters, and add a perturbation either to the gas density (cool core) or mm flux (radio source). Simulated observations of these mock clusters with the South Pole Telescope are generated, including realistic noise and background, yielding a realistic SZ signal-to-noise measurement.  

\subsection{Constructing Mock Galaxy Clusters}\label{mod}
	Since we are mostly interested in the SZ signal of our mock clusters, the key ingredient in our cluster models is the pressure profile $P(r)$. However, we must also model the cluster density profiles $\rho(r)$ since we wish to calculate the change in the pressure profile $\delta P$ due to a density perturbation $\delta \rho$, which in general is not $\delta P \propto \delta \rho$ as the temperature profile also shifts $T \rightarrow T - \delta T$. 
	Consequently, we need an additional constraint on our clusters to calculate $\delta P$, which we take to be hydrostatic equilibrium: $\frac{\mathrm{d}P}{\mathrm{d}r}  = -\rho \frac{\mathrm{d} \Phi}{\mathrm{d}r},$ where $\Phi$ is gravitational potential. Since there is no evidence that the underlying dark matter distribution is affected by the baryonic processes driving cool cores \citep{blanch13}, $\delta \Phi$ is straightforward to calculate, and using $P(r_{500})$ as a boundary value, it is possible to calculate $P + \delta P$.

	The details of our density and pressure parameterizations are now presented. For each non-cool core cluster, we draw a cluster mass $M_{500}$ and a redshift from uniform distributions (see Table \ref{table:dist}), with ranges motivated by the observed ranges in the SPT 2500 deg$^2$ survey \citep{bleem14}. The gas density profiles were then modeled by a slightly modified version of the functional form of \citet{V2006}:
	\begin{equation}
		\label{densparam}
		n_p n_e = \frac{n_0^2(r/r_c + \delta)^{-\alpha}}{\left[1+(r/r_c)^2\right]^{3\beta - \alpha/2}}\frac{1}{\left[1+(r/r_s)^3\right]^{\epsilon/3}},
	\end{equation}
	where the various length scales and slopes were sampled from realistic ranges for massive galaxy clusters by a Monte Carlo process, which we detail in Table \ref{table:dist}. For this work, we have ``regulated" the parameterization of \citet{V2006} by inserting a small factor $\delta = 0.1$ to obtain a finite density at $r=0$. This is necessary as we will eventually extract an SZ signal from the entire cluster, whereas the behavior of the profile very near $r=0$ is not of interest to observational works due to the limited spatial resolution of any survey.

	For a pure non-cool core, $\alpha = 0$ and equation (\ref{densparam}) reduces to the simpler form
	\begin{equation}
		n_p n_e = \frac{n_0^2}{\left[1+(r/r_c)^2\right]^{3\beta}}\frac{1}{\left[1+(r/r_s)^3\right]^{\epsilon/3}}.
	\end{equation}
	where $\alpha$, $\beta$, and $\epsilon$ are slope parameters for the $r \ll r_c$, $r_c \lesssim r \lesssim r_s$ and $r\gg r_s$ radial regimes. We convert $n_p n_e$ to the gas mass density via the relation $\rho = m_p n_e A/Z$ where $Z=1.199$ is the average nuclear charge, which is greater than unity because of ionized elements heavier than hydrogen, $A=1.397$ is the average nuclear mass, which is greater than $Z$ due to the presence of neutrons, and $n_e = Z n_p$ is the electron density in terms of the positive nuclei density $n_p$. The density normalization $n_0$ was chosen to enforce an average gas fraction $f_\mathrm{gas} = 0.125$ within $r_{500}$.

	For the unperturbed, non-cool core clusters, we adopt the simple temperature profile  

	\begin{equation}
		\frac{T(r)}{T_{0}} = 1.35\left[1+\left(\frac{r}{0.6 \, r_{500}}\right)^2\right]^{-0.45},
	\end{equation}
	which is the ``universal" temperature profile of \citet{V2006} with the term accounting for the central temperature drop suppressed.

	For each non-cool core cluster, we generate 14 progressively cuspier cool-core clusters. We start by duplicating each non-cool core cluster and shifting $\rho \rightarrow \rho + \delta \rho$. To avoid adding more free parameters (a cooling radius, etc.) which may be correlated with the other model parameters, we simply increment $\alpha$ to steepen the density profile towards the center of the cluster. Each successive cluster has $\alpha_n = \alpha_{n-1} + 0.25$. The density normalization $n_0$ was also reduced by an appropriate factor to keep the gas mass within $r_{200}$ constant.

	Finally, the assumption of hydrostatic equilibrium was used to recalculate the pressure profile of the cool core cluster $P \rightarrow P+\delta P$:
	\begin{equation}
		P+\delta P = \int \mathrm{d}\tilde{r} \left [ \frac{\rho+\delta \rho}{\rho}\frac{\mathrm{d}P}{\mathrm{d}\tilde{r}}-\frac{(\rho+\delta \rho) G\,\delta M}{\tilde{r}^2} \right ].
	\end{equation}
	We vary $P(r_{200})$ until the ``internal energy" $\propto \int P \mathrm{d}V$ of the cool core cluster out to $r_{200}$ agrees with its non-cool core counterpart. Note that in general, this means that $\delta P(r_{200}) \ne 0$, though at $r_{200}$, $\delta P \ll P$. 
	
	The above methods were used to generate $200\times 15$  mock clusters which ranged from pure non-cool core $\alpha = 0$ to strong cool core $\alpha = 3.5$. We chose the maximum value $\alpha = 3.5$ as even strong cool core clusters like Phoenix have $\alpha < 3.5$.
	
	\ctable[star, caption = {Parameters. All radii are in units of kpc. We report $1\sigma$ uncertainties.}, label = {table:dist}]
	{llll}	{
			\tnote[a]{We fit triangle distributions to the empirical distributions derived from \citet{new} to avoid the long Gaussian tails that lead to unphysical clusters. We report the (min, max, mode) of the distribution in the range column.}
			\tnote[b]{The mock SZ signal is virtually $\epsilon$-independent. We choose $\epsilon$ to let $\mathrm{d}\log P/\mathrm{d}\log r = - 0.90$ at large radii, as given by the universal temperature profile of \citet{upress}.}
			}
		{ \FL
		Parameter & Range & Distribution &	Notes\ML
		$M_{500}/M_\odot$ 	& $[1\! \times\! 10^{14}, 2.1\! \times \!10^{15}]$ 	& Log-Uniform	& SPT 2500 deg$^2$; \citet{bleem14}\NN
		$z$				 	& $[0,2]$ 						& Uniform			& SPT 2500 deg$^2$; \citet{bleem14}\ML 
		$\beta$ 			& $(0.28, 0.88, 0.71)$			& Triangle\tmark[a]	& \citet{new},\citet{V2006}\NN
		$\epsilon$ 			& $7.1 - 6 \beta$\tmark[b]		& ---				& \citet{and2011}\NN
		$r_c/r_{500}$		& $(0.067, 0.26, 0.47)$			& Triangle\tmark[a]	& \citet{new}\NN
		$\log r_s/r_{500}$ 	& $1.21 $ 						& ---				& \citet{new}\LL
		}

\subsection{Model Validation}\label{add}

\begin{figure*}[t]
	\centering
	\subfigure{
	\includegraphics[scale=0.45, trim = 2cm 13cm 2.cm 3cm, clip=true]{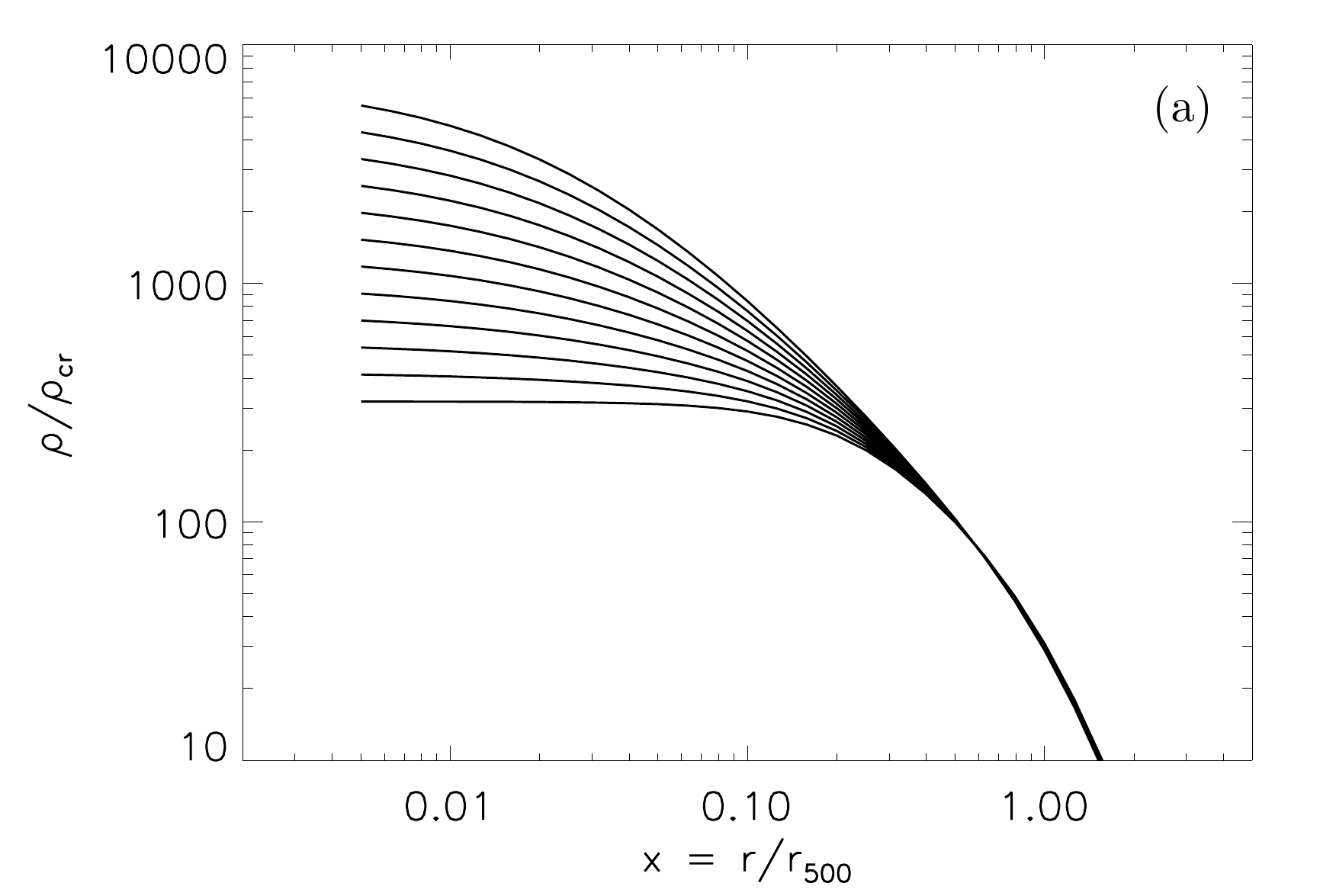}
	}
	\subfigure{
	\includegraphics[scale=0.45, trim = 2cm 13cm 2.cm 3cm, clip=true]{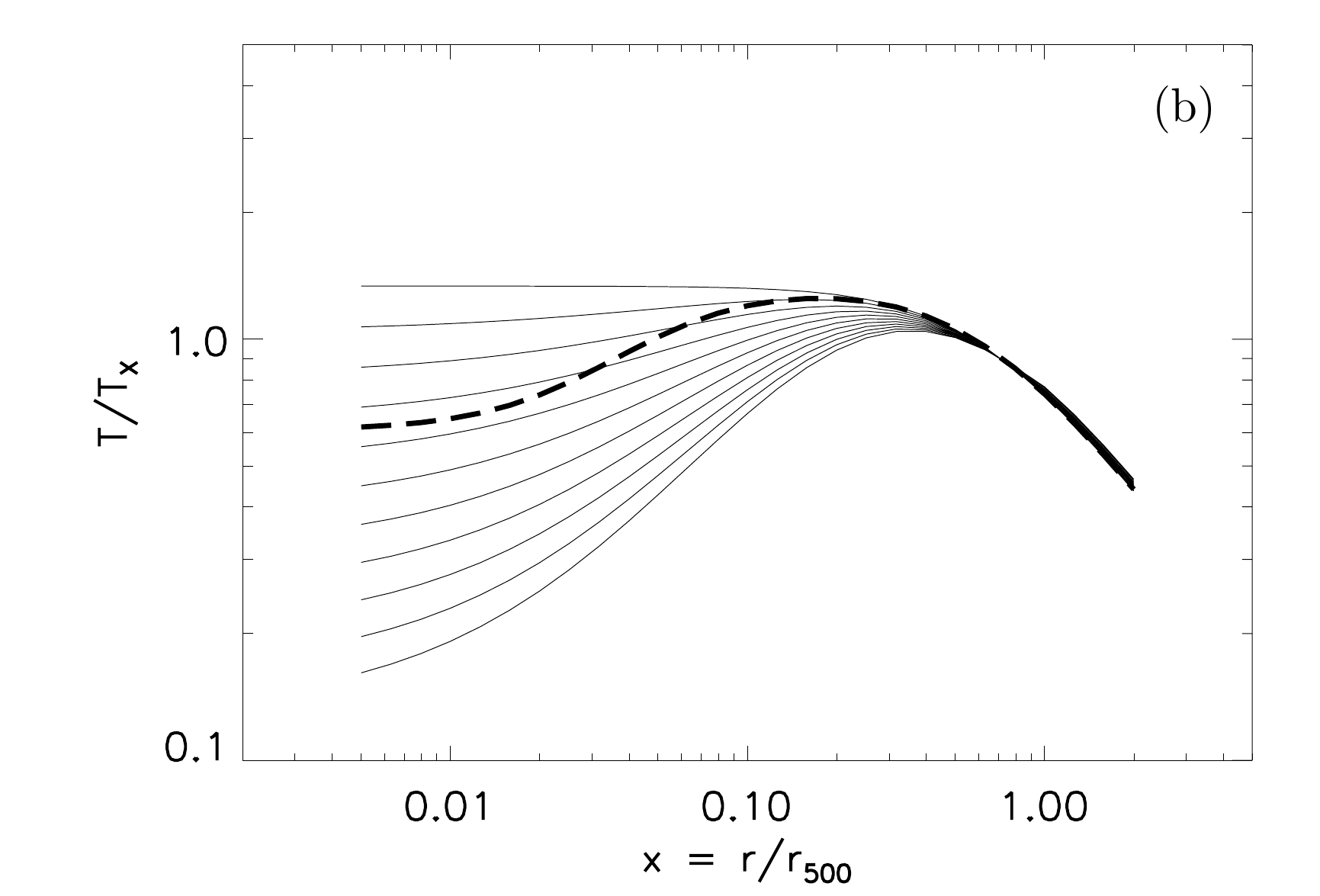}
	\label{fig:temp}
	}
	\subfigure{
	\includegraphics[scale=0.45, trim = 2cm 13cm 2.cm 3cm, clip=true]{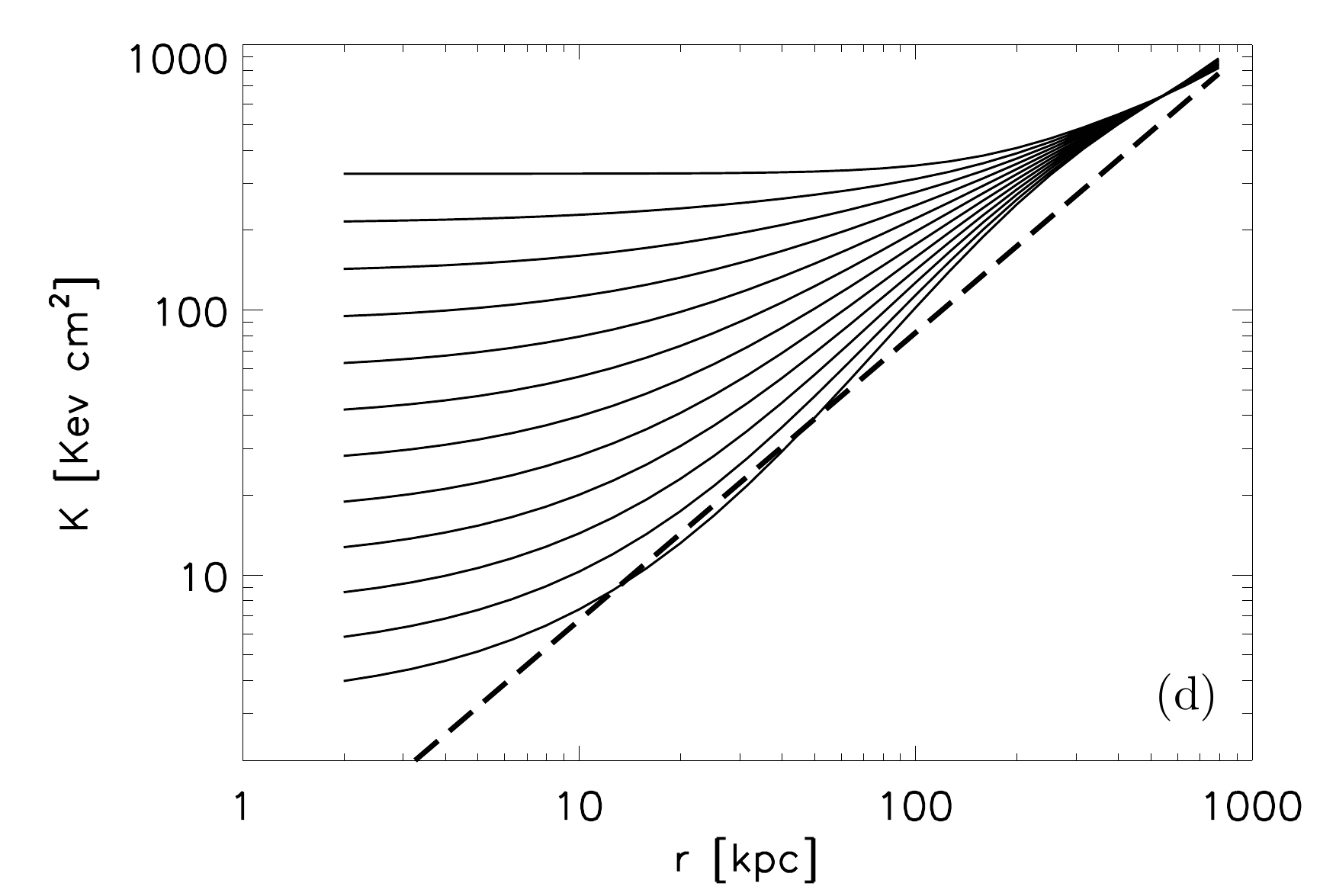}
	\label{fig:entropy}
	}	
	\subfigure{
	\includegraphics[scale=0.45, trim = 2cm 13cm 2.cm 3cm, clip=true]{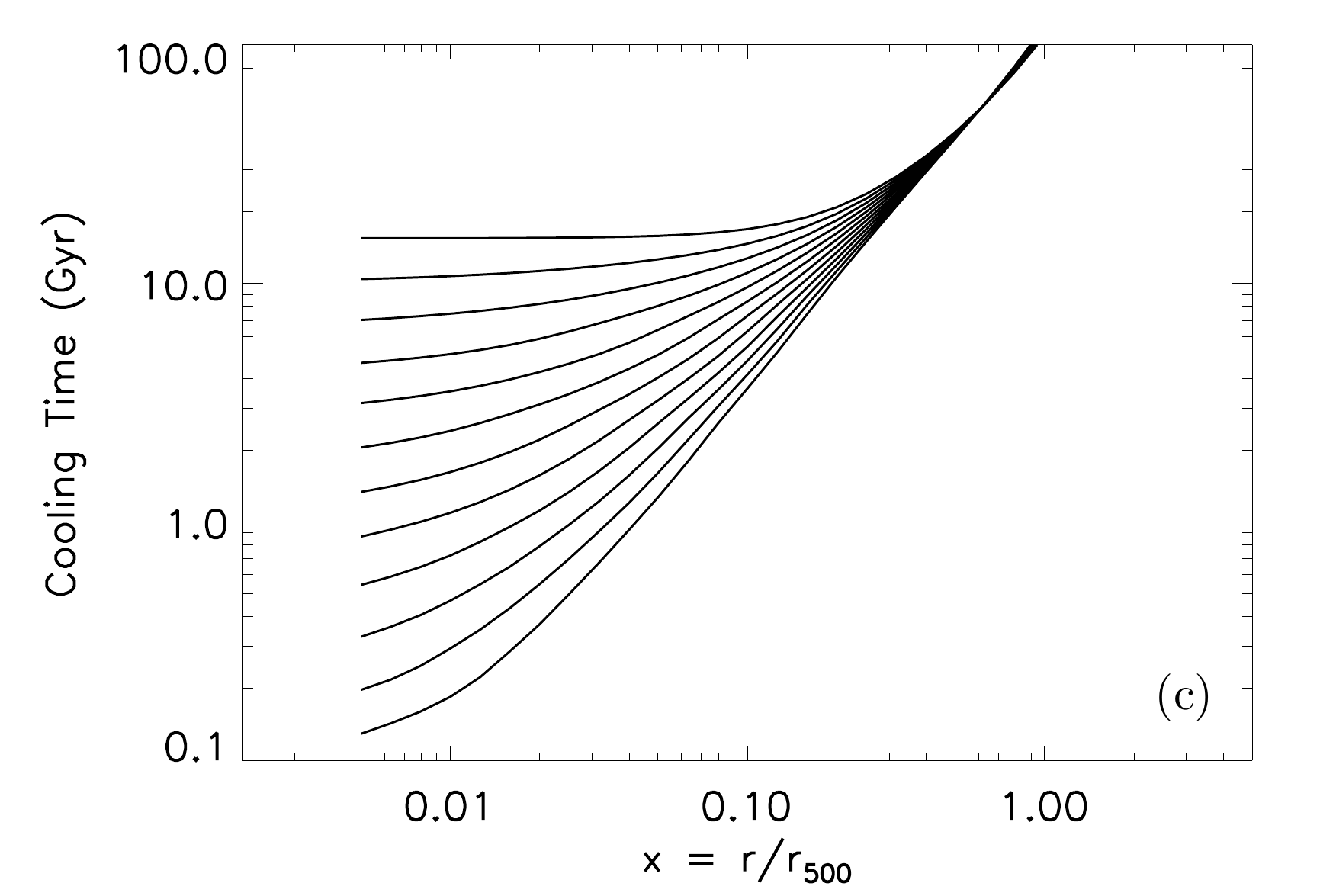}
	\label{tcool}
	}	
	\caption{Clockwise from the upper left: density, temperature ($T_x  = \int n^2 T \, \text{d}V/\int n^2 \, dV$), cooling time, and entropy as a function of cluster radius for a typical mock cluster. All of the displayed profiles were obtained by cloning a single non-cool core core cluster and adding density perturbations. These plots represent $1/200$th of the total dataset. In the upper right plot, the thick dashed line represent the universal temperature profile of \citet{V2006}. In the lower left plot, the thick dashed line represents the predicted entropy profile slope from non-radiative simulations of \citet{voit05}.
	\label{Fig:profiles}}
\end{figure*}

	Additional quantities were derived to demonstrate how the wide range of observed cluster properties are captured by our methodology.

	\begin{enumerate}
	
		\item {\it Temperature.}
			Our derived temperature profiles are displayed in Figure \ref{fig:temp}. The cool core profiles exhibit a central temperature drop similar to the average cool core profiles observed in X-ray selected samples \citep{V2006}. The characteristic temperatures of our clusters range from $\sim\!1$ keV to $\sim\!30$ keV, which is in agreement with observations of massive clusters. Furthermore, $T_\text{min}/T_0$, which corresponds to the ratio of the central cool core core temperature to the $\alpha=0$ central temperature, range from $0.1$ to $0.85$ in the \citep{V2006} clusters, which is completely covered by our simulated clusters. Similarly, our simulations cover the range of temperature profiles observed in \citep{mcdonaldstate}, which find that the fractional temperature in the inner $0.04 R_{500}$ is $T_\text{core}/T_{500} = 0.74^{0.09}_{-0.04}$ for high redshift, cool core clusters and higher for low redshift cool core clusters.
		
		\item {\it Entropy Parameter.}
			The entropy profiles of our mock clusters are derived in an effort to show their resemblance to observed clusters. 
			Following the X-ray survey conventions, an entropy parameter $K \equiv k_B T n_e^{-2/3}$ is introduced. 
			We plot this parameter as a function of radius in Figure \ref{fig:entropy}. Consistent with \citet{entr}, \citet{Hudson:2010} and \citet{new}, the cool core clusters have a central entropy $K_0 \lesssim 100 \, \mathrm{keV\,cm^2}$, whereas for non-cool core clusters $K_0 \gtrsim 100 \, \mathrm{keV\,cm^2}$. The slope of our entropy profiles outside of the core is also in reasonable agreement with the accretion shock model of \citet{voit05}, which represents the ``baseline" entropy profile of clusters if radiative and other non-gravitational processes are ignored.
		
		\item {\it Cooling Time.} 
			The cooling time $t_\mathrm{cool}$ has been shown by \citet{Hudson:2010} to be a quantity that can be used to segregate cool cores from non-cool cores. In Figure \ref{tcool}, $t_\mathrm{cool} = 3 kT/\left(n_e\Lambda\right)$ is plotted as a function of radius. Here, $\Lambda(T,Z)$ is the cooling function given by \citet{coolingfun}. Most non-cool cores have central cooling times $t_{\mathrm{cool},0} \gtrsim 1/H_0$ whereas the cool core clusters have cooling times $t_{\mathrm{cool},0} \sim 1 \mathrm{Gyr}$, consistent with the findings of \citet{Hudson:2010}.
		
	\end{enumerate}

\subsection{Sunyaev Zel'dovich Maps}\label{sz}

	For each mock cluster, we constructed an SZ map with an angular resolution of $\Delta \theta = 0.125 \, \mathrm{arcmin}$, which was converted to spatial resolution $\Delta r_\mathrm{2D} = d_A(z)\Delta \theta $ where $d_A(z)$ is the angular diameter distance to the cluster. The SZ line of sight integral was then calculated at each pixel:

	\begin{equation}\label{eqn:sz}
	\Delta T/T_\mathrm{CMB} = \int_{\vec{\theta}} \mathrm{d}l \left\{ f_\mathrm{sz}(\nu,T) \frac{\sigma_t n_e k_B  T}{m_e c^2} \right \}	\end{equation}

	\begin{equation}
	f_\mathrm{sz}(\nu, T) = \left(X \frac{e^X +1}{e^X -1} -4\right)\left[1+\delta_\mathrm{sz}(X, T)\right]
	\end{equation}
	where $X\equiv h\nu / k_B  T_\mathrm{CMB}$, $\sigma_t$ is the Thomson cross section, $m_e$ is the mass of the electron, $c$ is the speed of light, and $\delta_\mathrm{sz}$ is the relativistic correction as given by \citet{itoh}. For our purposes, it sufficed to evaluate $\delta_\mathrm{sz}(T)$ at $T(r_{500})$.
	We made simulated SZ maps at 97.6 and 152.9 GHz, corresponding to the effective frequency of the SPT observing bands used for cluster-finding \citep{sptintro}. Our mock SZ maps were generated at a resolution $\ll 1'$ and then later convolved with the SPT beam which has an effective FWHM of $\sim 1.2'$. 

	The mock SZ maps were then spatially filtered in a way identical to the SPT cluster-finding algorithm used in \citet{spt13}.  A multi-frequency matched spatial filter \citep{tegmark,melin} was applied to the SZ maps in Fourier space, which accounted for the cluster gas profile, and the other sources of astrophysical signal and instrumental noise expected in the SPT maps.  The spatially filtered signal at the true cluster position was then re-normalized so that it was equivalent to the unbiased SPT significance, $\zeta$, used in \citet{vander10}, which would effectively correspond to the signal-to-noise of cluster in a SPT map.
	
	These SZ maps were run through the SPT cluster-finding pipeline, which adds noise in the form of point sources and CMB anisotropy maps, before attempting to detect clusters. For each cluster, the SPT pipeline then reported $\zeta$, the ``unbiased significance" \citep[see][]{vander10, sptcosmo} that roughly corresponds to the cluster signal-to-noise ratio.

\subsection{Active Galactic Nuclei}\label{AGN}
	In order to test the effects of including radio-loud AGN, we created mock clusters with the following grid of parameters:
	\begin{itemize}
		\item $M_{500}/2 \times 10^{14} M_\odot = 10^{0}, 10^{1/2}, 10^{1}$
		\item $z = 0.3, 0.4, 0.5, 0.7, 1.1, 1.4, 1.7$
	\end{itemize}
	and we include a radio-loud AGN at the cluster center, with a range of luminosities:
	\begin{itemize}
		\item $\log L_{1.4} = 23$ -- $29$, $\Delta \log L_{1.4} = 1.0$
	\end{itemize}
	where $L_{1.4}$ is the 1.4 GHz luminosity in $\mathrm{W}\,\mathrm{Hz}^{-1}$. All other parameters are set to their median values. The range in $\log L_{1.4}$ covers the {most luminous brightest cluster galaxy (BCG) in the 152 X-ray BCG sample of \citet[]{Sun09}}, allowing for unusually shallow spectrum systems that would generate a large bias. 
	
	We convert these luminosities to the SPT observing frequencies by assuming a spectral slope of $\alpha_s = 0.89$, where flux scales with $\nu^{\alpha_s}$.  This slope was the median spectrum between 1.4 and 30 GHz for radio galaxies in a sample of 45 massive clusters observed by \citet{sayers}, however it is always possible to reinterpret the results for different choices slopes. We discuss deviations from this assumption in Section 3.1. As a control, a point source is not added to one cool-core/non-cool core cluster at each mass and redshift.

\section{RESULTS}

\begin{figure*}
	\centering
	\subfigure{
	\includegraphics[scale=0.42, trim = 1.4cm 0 0 0, clip=true]{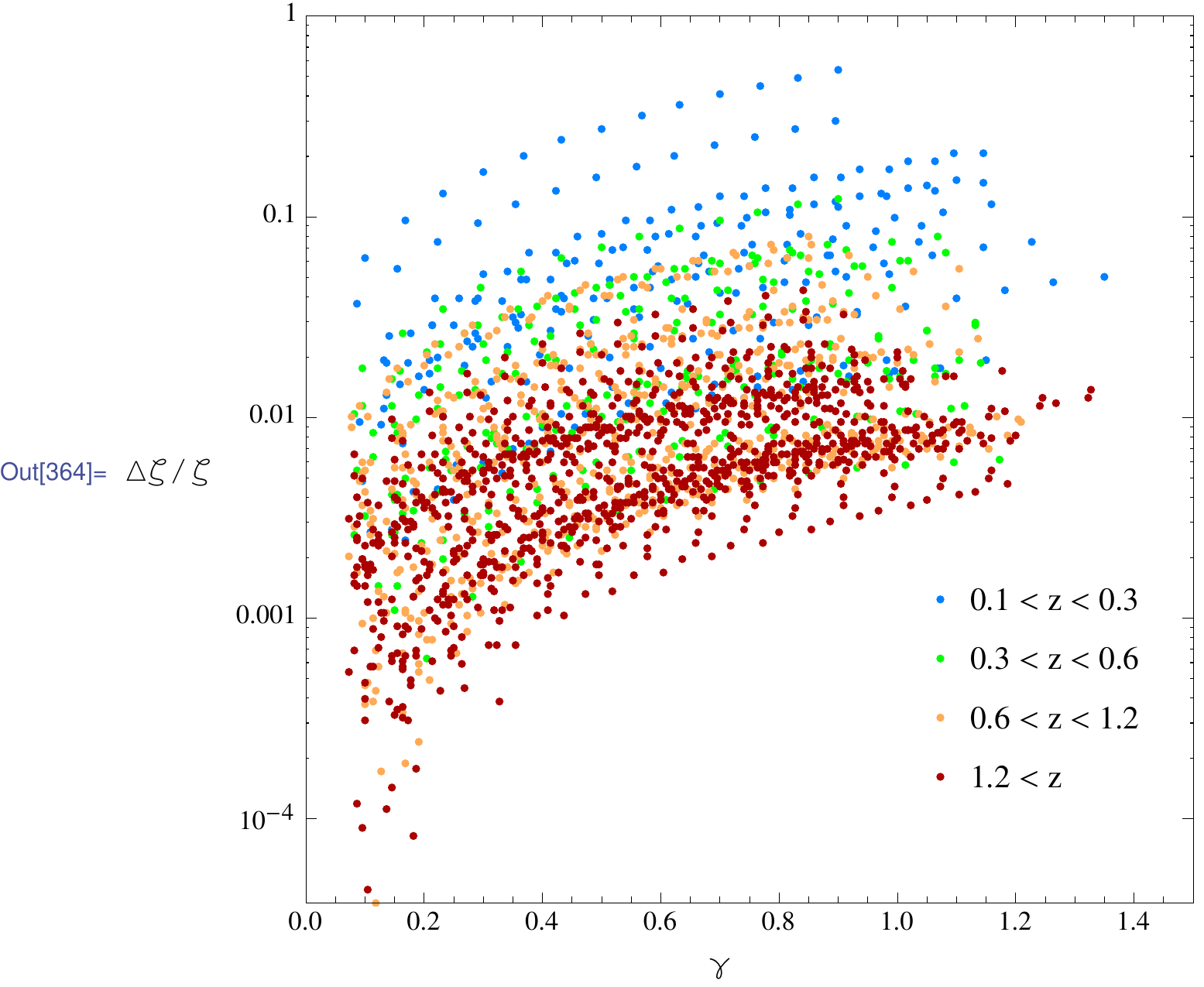}
	}
	\subfigure{
	\includegraphics[scale=0.42, trim = 3cm 0 0 0, clip=true]{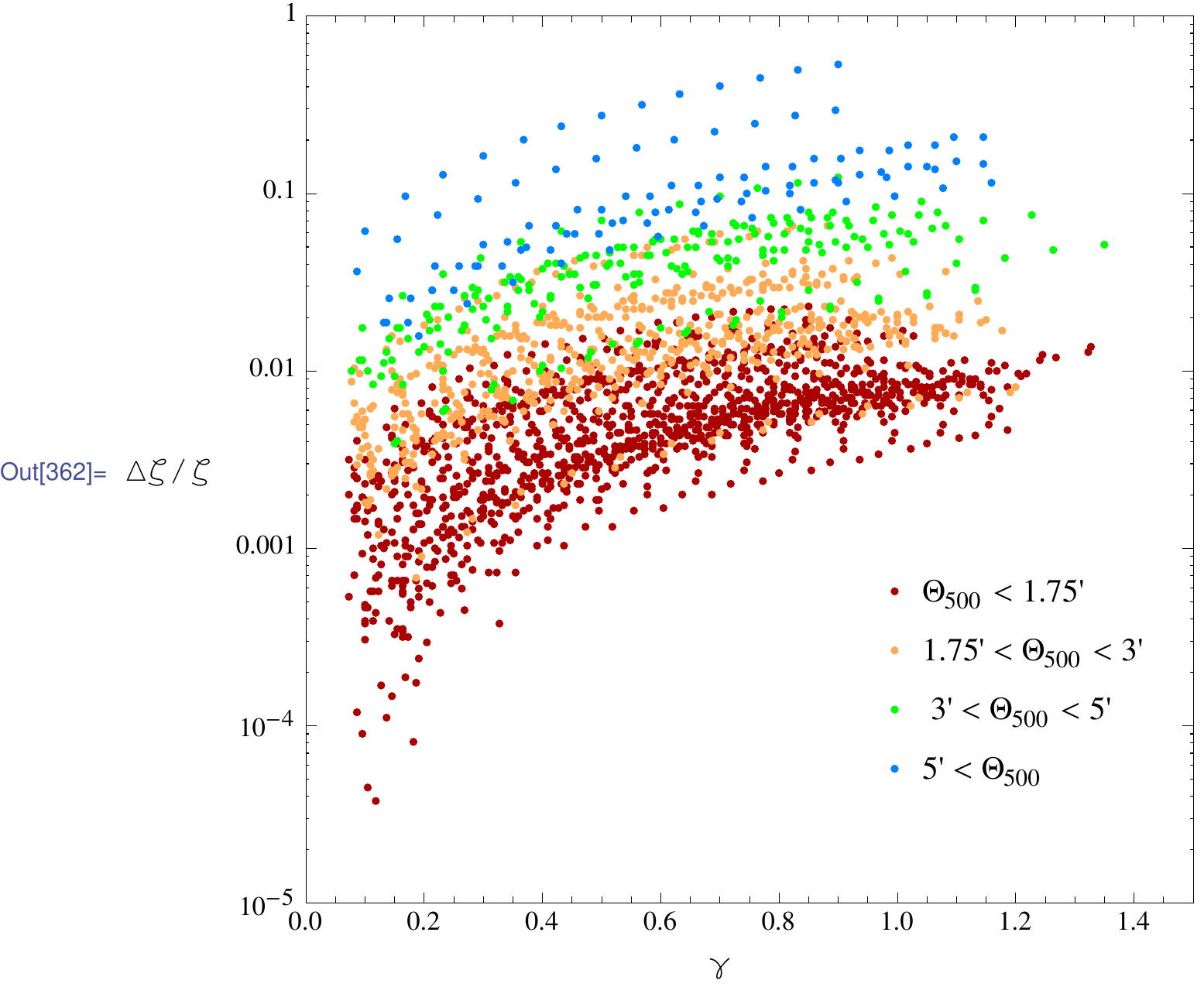}
	}
	\subfigure{
	\includegraphics[scale=0.42, trim = 3cm 0 0 0, clip=true]{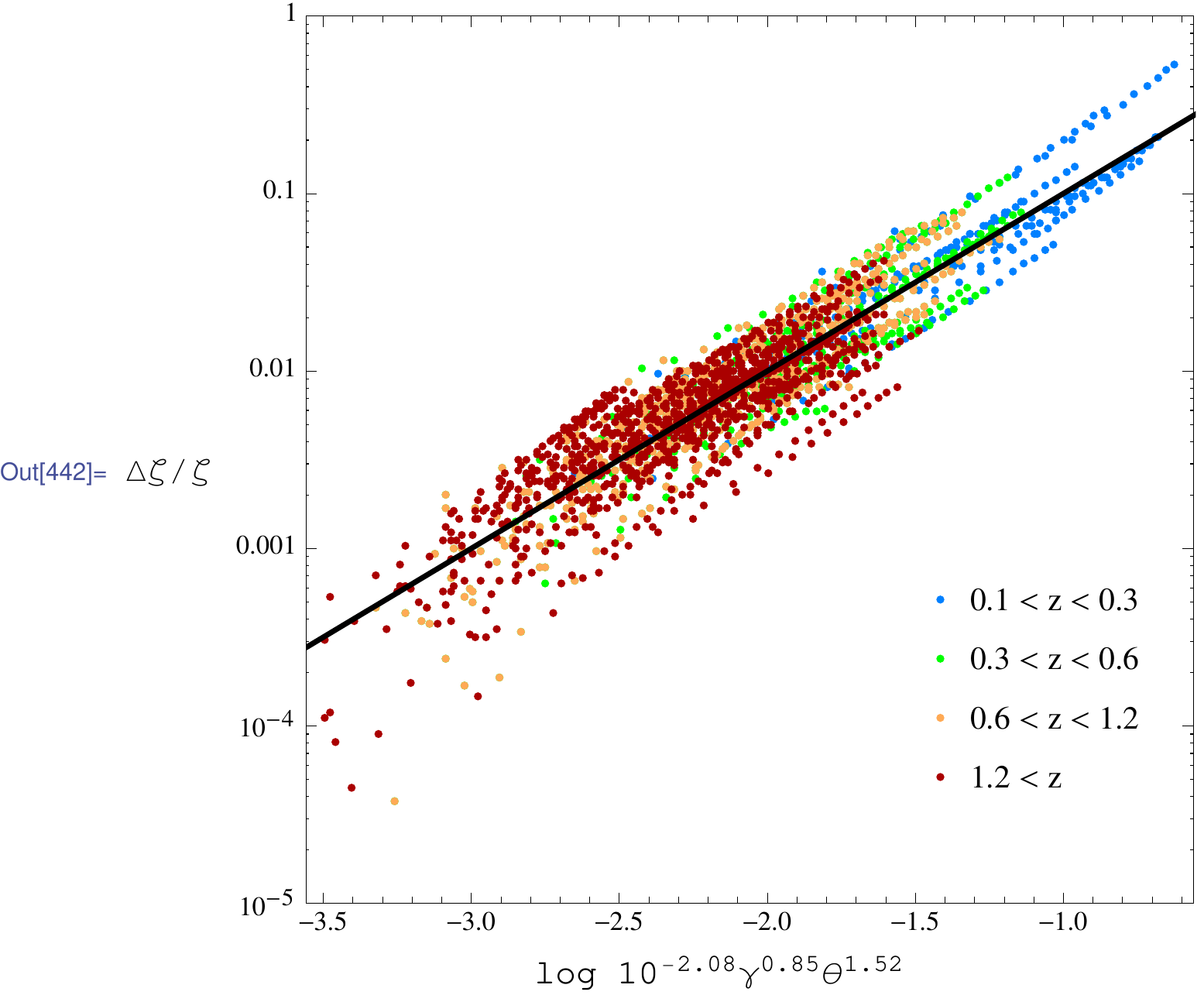}
	}
	\caption{Dependence of the SZ bias on the cuspiness ($\gamma \equiv - \frac{\mathrm{d} \log \rho}{\mathrm{d}\log r}$), redshift ($z$), and angular size ($\Theta_{500} \equiv r_{500}/d_A(z)$) for 3000 mock clusters. In the left panel, the point color corresponds to redshift of the mock cluster. In the central panel, we color by the angular size of the cluster. The clear separation into colored bands in this plot suggests that the bias is more fundamentally tied to the angular size than it is to the redshift. In the right-most panel, we show the edge-on projection of the best-fit three-dimensional plane (Equation 7). The $\sim$0.5 dex scatter is primarily due to variations in the cluster gas density profile, and the imperfect classification of cool core strength based on the $\gamma$ parameter.}
\label{fig:main}\end{figure*}

\begin{figure*}
	\centering
	\includegraphics[scale = 1.1, trim = 0cm 0 0 0, clip = true]{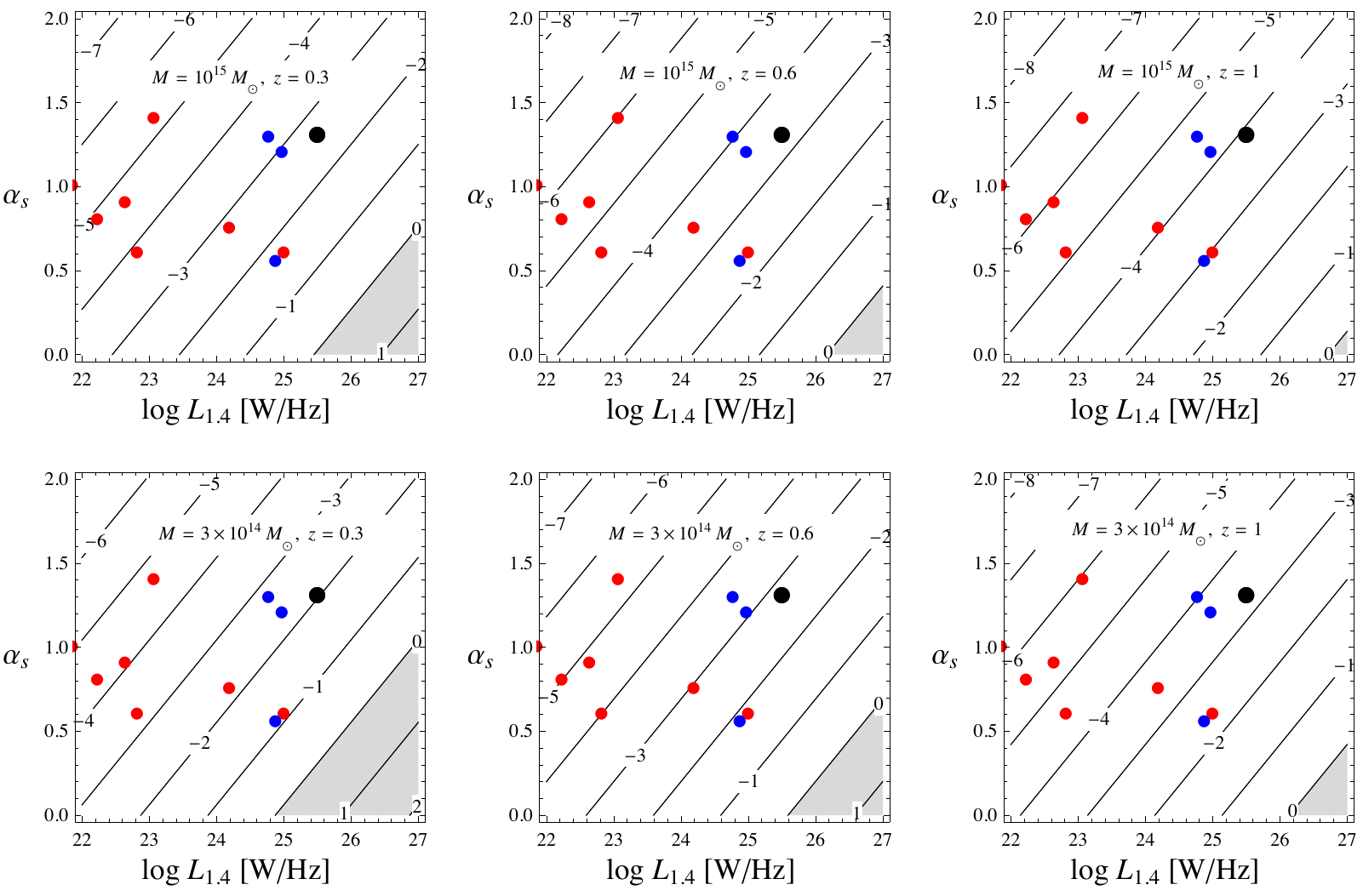}
	\caption{Dependency of $\Delta\zeta/\zeta$ on $M_{500}$, redshift, radio luminosity, and spectral slope. The contours represent lines of constant $\log ( \Delta \zeta/\zeta)$. If an AGN exists in the shaded region, it will overpower the SZ signal. The large black dot represents a Phoenix-like cluster at multiple redshifts. Even at $z=0.3$, a Phoenix-like cluster will exhibit only a $1\%$ bias. The most extreme systems exhibit a $\sim 10\%$ bias at $z \sim 0.3$. Red dots represent a sample of X-ray selected massive elliptical galaxies from \citet{dunnr} and blue dots represent extreme AGN hosted in massive clusters \citet{MACS}.
	\label{fig:AGN}
	} 
\end{figure*}
	
\begin{figure}
	\centering
	\includegraphics[scale = 1.2, trim= 0cm 0cm 0cm 0cm, clip=true]{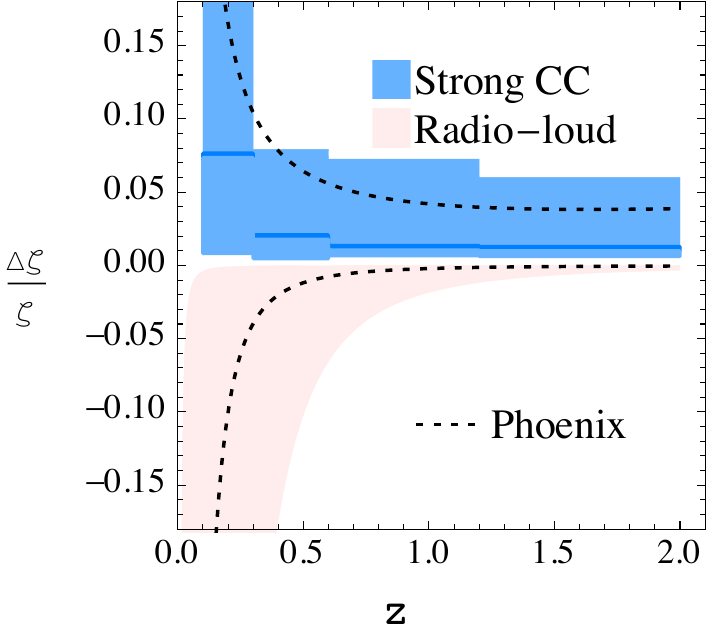}
	\caption{Strong cool core $(\alpha = 2.75)$ bias and radio-loud AGN bias. The black, dotted lines represent the massive cool core ``Phoenix'' cluster, with radio spectral slope $\alpha_s = 1.3$ \citep{mcdonaldstate} and X-ray properties given in \cite{new}, for a range of redshifts. 
	The pink shaded region represents a series of radio-loud AGN with spectral slope $\alpha_s \ge 0$, which demonstrates the range in possible AGN bias. The dark blue lines are the median bias in each bin and the light blue regions represent $2 \sigma$ confidence intervals.
	In general, the cool core bias and the radio-loud AGN bias work to roughly cancel each other.\label{fig:cc-z}}
\end{figure}


	In Figure \ref{fig:main}, we show the fractional change in the SPT detection significance as a function of the cool core cuspiness \citep{V2007}, angular size ($\Theta_{500} \equiv r_{500}/{d_A(z)}$), and redshift. At $\Theta_{500} < 3'$, the SPT detection significance changes by $\Delta \zeta /\zeta = (\zeta_\text{CC} - \zeta_\text{NCC})/\zeta_\text{NCC} \lesssim 0.1$ for any type of cool core at any redshift. The cool core bias is most pronounced for low-redshift, high mass systems, where the angular size of the SPT beam is much smaller than the angular size of the cluster so that the inner pressure profile becomes important. 
	In principle, the resulting redshift-dependence in the bias might lead to a $\sim 2\%$ bias towards cool core clusters from $z \sim 1.5$ to $z\sim 0.3$ in the observed cool core fraction $CCF \equiv N_\text{cool core}/N_\text{clusters}$ evolution, but these effects should be negligible when compared to the $\sim 30\%$ evolution in $CCF$ reported in e.g. \citet{new}.

	In real observational studies it may be difficult to determine the density shape parameter $\alpha$, as this requires a multi-parameter fit to the $n_p n_e$ profile and assumptions on the shape of the cool core/non-cool core profiles. To ameliorate the situation, we calculate the ``cuspiness" parameter $\gamma \equiv -\mathrm{d} \log \rho / \mathrm{d} \log r|_{r=0.04 r_{500}}$ \citep{V2007} of all simulated clusters.
	Since this quantity does not refer to the details of our parameterizations, it serves as a model-independent indicator of cool core strength; correlations between these quantities and $\Delta\zeta/\zeta$ are shown in Figure \ref{fig:main}. 
	The bias is well-captured by the following relation:
	\begin{equation}
		\langle \delta \zeta/\zeta \rangle = 10^{-2.08 \pm 0.01} (\Delta \gamma)^{0.85 \pm 0.01}\left( \frac{\Theta_{500}(z)}	{\mathrm{arcmin}}\right)^{1.52\pm 0.02}
		\label{big}
	\end{equation}
	where $\Delta \gamma = \gamma - \gamma|_{\alpha=0} \approx \gamma-0.1$.
	This expression gives a practical method for estimating the SZ bias of a given cool core cluster, assuming that $z > 0.1$. 

\subsection{Radio bias}

	{Assuming a typical spectral slope of $\alpha_s = 0.89$ \citep{sayers}, only sources with $\log L_{1.4} > 26$ result in $|\Delta \zeta/\zeta| \gg 10 \%$ for $z > 0.3$ and $M > 3 \times 10^{14}$. Consequently, none of the 152 radio-loud AGN listed in \citet{Sun09} would produce a bias in excess of $10\%$ if located in massive clusters. 	More generally, the radio-loud point source bias can be readily estimated by the following fit to our mock clusters (see \S2.4): 
	\begin{equation}
		\label{eqn:agn}
		\langle \delta \zeta/\zeta \rangle = -0.03 \left(\frac{\nu_\mathrm{SZ}}{1.4 \mathrm{GHz}} \right)^{-\alpha_s} \left(\frac{S_{1.4}}{\mathrm{mJy}}\right) \left(\frac{M_{500}}{10^{14} M_\odot}\right)^{-1}
	\end{equation}

	In Figure \ref{fig:AGN} we show a contour plot of $\Delta \zeta/\zeta$ on $\Theta_{500}(M_{500}, z), \alpha, L_{1.4}$ parameter space. We display flux and luminosity data from a sample of X-ray selected massive elliptical galaxies from \citet{dunnr} and extreme AGN hosted in massive clusters \citet{MACS}.
	As can be seen by Figure \ref{fig:AGN}, the fractional change in SPT significance will be strongly dependent on the assumed value of $\alpha_s$, which has a fairly large scatter. For example, \citet{coble} measures a median value of $\alpha_s = 0.72$ between 1.4 and 28.5 GHz. \citet{lin09} explicitly measure the radio emission at frequencies ranging from 1.4 to 43 GHz in an X-ray selected sample of clusters and find mostly steep spectra ($\alpha_s > 0.5$), but also find a substantial number of flat or even inverted spectra. Neither of these studies -- or any published studies of radio galaxies in clusters, for that matter -- extend to $\sim$100\,GHz, so there is still substantial uncertainty in the value of $\alpha_s$ that we should be using.
	
	
	In Figure \ref{fig:cc-z} we show the strong cool core $(\alpha = 2.75)$ bias and the radio-loud AGN bias as a function of redshift. Interestingly, while SZ surveys are biased \emph{towards} cool cores, they are biased \emph{against} radio-loud AGN, and this bias is of the same order of magnitude as the strong cool core bias, defined as the average $\Delta \zeta /\zeta$ over cool cores with $\alpha = 2.75$. Since powerful radio galaxies tend to live in the center of strongly-cooling galaxy clusters, the net bias is partially canceled (see figure \ref{fig:cc-z}). {In principle, low-redshift galaxy clusters that harbor a powerful, radio galaxy ($\log L_{1.4} \gtrsim 25$) but lack a cool core will be under-represented in SZ surveys. However in practice, such systems appear to be uncommon in nature (see e.g. \citet{Sun09} and \citet{dunnr}).} The AGN bias is more strongly redshift dependent than the cool core bias due to the $\sim 1/d_L^2$ flux dimming where $d_L$ is the luminosity distance, whereas the evolution in the simulated cool core bias is a result of SPT beam effects and weak dependencies of our parameterizations on $\rho_\mathrm{cr}(z)$. 
	
\section{Robustness of simulations}
In this section we estimate how relaxing several of our simplifying assumptions could change our bias estimates. Though a thorough treatment of each of these issues is beyond the scope of this paper (and should be addressed by, e.g. N-body simulations), an estimate on the relative importance of these effects can be obtained by recasting these effects into an $M_{500}$ dependence.

\subsection{Deviations from hydrostatic equilibrium and non-thermal pressure}
In the absence of complicated astrophysics, a galaxy cluster will equilibrate on dynamical timescales $t_\text{dynamic} \sim R_\text{500}/v \sim (G\rho)^{-1/2} \sim 10^9$ yr. For hydrostatic equilibrium to generically hold, the cooling timescale should be much longer than the dynamical timescale $t_\text{cool} \gg t_\text{dynamic}$ so that hydrostatic equilibrium will be restored efficiently when gas cools. 
If the cooling timescale is shorter than the equilibrium timescale in the core of the cluster, we would expect pressure to be lower than what is required to maintain hydrostatic equilibrium, since gas must be sinking towards the core. This would mean that the bias should be generically lower than what we have derived, since the observed change in pressure should be smaller than that required to maintain hydrostatic equilibrium $\delta P_\text{obs} \le \delta P_\text{eq}$. 

Throughout our discussion, we have assume that pressure is purely thermal, e.g. $P = n kT$. Analytic work \citep{shi}, numerical simulations \citep{dolag, sembo13, nelson14,battaglia14}, and multi-wavelength observations \citep{planck13, linden14, donahue, sereno14}, however, have shown that non-thermal pressure can contribute significantly ($\sim 20\%$; increasing with cluster radius), which can lead to an underestimate of $M_{500}$ by about $\sim 10\%$ \citep{shi}. Sources of non-thermal pressure include turbulence from intracluster shocks, magnetic fields, and cosmic rays. If turbulence dominates the non-thermal pressure, non-thermal pressure will persist for on roughly the dynamical timescale \citep{shi}.

From equation (8) we see that bias scales as $1/M$, so to first order, we expect that adding non-thermal pressure will rescale the bias by $(1 + P_\text{non-therm}/P_\text{therm})^{-1}$, if we assume that the processes responsible for forming cool cores and providing non-thermal pressure are uncorrelated. As clusters evolve, this term could introduce an additional redshift dependence on the cool core bias, if the processes generating non-thermal pressure evolve with redshift.

\subsection{Varying $f_\text{gas}$}

Throughout this work, we have assumed that $f_\text{gas} = 0.125$ with no scatter. However, the numerical simulations of \citet{battag} which included radiative physics and AGN feedback estimate a scatter $\sim 10\%$ in $f_\text{gas}$ at $r < r_{500}$. \citet{eck13} suggest that cool core clusters and non-cool core clusters differ in $f_\text{gas}$ by a few percent, with cool core clusters more reliably tracing the cosmic baryon fraction.

For a fixed number of baryons, a lower $f_\text{gas}$ will result in an increase in the pressure normalization, in order to overcome the steeper potential due to additional dark matter. For a cluster in hydrostatic equilibrium, $kT \propto M/r$ giving $P \propto M^{2/3} \sim (1/f_\text{gas} )^{2/3} M_\text{baryon}^{2/3}$, which means that the effects of changing $1/f_\text{gas}$ will be roughly equivalent to changing $M_\text{500}$. Equations (7) and (8) can therefore be rescaled by a factor $(f_\text{gas}/0.125)$. Adding intrinsic scatter in $f_\text{gas}$ would then simply add scatter in the SZ bias. If $f_\text{gas}$ varied with redshift, this could lead to a further evolutionary factor in Equation (8). However, \citet{battag} found little evidence for such a redshift dependence, as has long been assumed \citep{white93}.

\subsection{Effects and Evolution of AGN}
We purposefully exclude any evolution in AGN properties as a function of cluster mass and redshift, despite evidence that such links exist \citep[e.g.,][]{ma13}. However, by casting Equation (8) in terms of the cluster mass, redshift, and radio luminosity, we allow the direct incorporation of such trends into the bias estimate. By fully sampling a grid of parameters (see \S2.4), we make certain that, regardless of how AGN evolve, we understand how they influence the SZ signal at all redshifts and cluster masses.

Radio-mode AGN feedback can modify the gas distribution of galaxy clusters as gas is heated and expelled from the core, leading to deviations from hydrostatic equilibrium in the inner region. Such processes will typically result in less pressure than predicted by hydrostatic equilibrium, since the gas is supported by feedback in addition to thermal pressure.

\section{Application to Well-Known Systems}
Using Equations 7 and 8, we can now calculate how biased SZ surveys are (or are not) for some well-studied extreme systems. First, we consider two of the strongest known cool cores: Abell~1835 \citep[$\gamma = 0.85$, M$_{500} \sim 10^{15}$ M$_{\odot}$;][]{a1835} and the Phoenix cluster \citep[$\gamma = 1.29$, M$_{500} \sim 1.3 \times 10^{15}$ M$_{\odot}$;][]{nature}.  Assuming both of these clusters are at $z=0.6$, we find $\Delta\zeta/\zeta = $3.6\%\ and 5.9\%, for Abell~1835 and Phoenix, respectively. At $z=1.0$, these biases are reduced to 2.8\%\ and 4.2\% (see also Figure \ref{fig:cc-z}). Given that these are two of the most extreme cool cores known, we expect the typical bias towards selecting cool cores in SZ surveys to be $\ll5\%$ at $z>0.5$.

Figure \ref{fig:AGN} quantifies the radio bias for a variety of nearby clusters from \cite{dunnr} and \cite{MACS}, but here we specifically look at two well-known nearby radio-loud central galaxies: Hydra A \citep[$S_{1.4} = 40.8$ Jy, $z=0.055$;][]{birzan04} and Cygnus A \citep[$S_{1.4} = 1600$ Jy; $z=0.056$;][]{birzan04}, the latter being one of the most powerful examples of radio-mode AGN feedback known. At $z=0.6$, these two clusters would be biased low in SZ significance by 1.8\%\ and 110\%, respectively. At $z=1.0$, these biases are further reduced to 0.5\%\ and 31\%. Thus, while SZ surveys may miss \emph{the most extreme radio-loud clusters}, this bias is rapidly reduced with increasing redshift. In nearby X-ray-selected clusters, $\ll$1\% of clusters have radio luminosities as high as Cygnus A (Hogan et al.\ in prep), so we expect this bias to be small overall.

\section{Mass Bias in SZ Surveys}

	To translate SZ observations to cosmological constraints, it is necessary to estimate mass distribution of an ensemble of clusters. To this end, \citet{vander10} and \citet{sptcosmo} have adopted $\zeta$ as a proxy for $M_{500}$ by assuming a scaling relation $\zeta \propto M^{B}$. 
	As a consequence, the true $M_{500}$ of cool core clusters will be underestimated by an amount on the order of $\int P(\Theta_{500}, \gamma) \langle \delta \zeta/\zeta \rangle d\Theta_{500} \, d\gamma$ on average (for $B\sim 1$) if the calibration was performed using only a sample of non-cool core clusters. In principle, this redshift-dependent bias should translate into a distorted $N(M,z)$; however the log-normal intrinsic scatter in $\zeta$ given mass has been measured to be $0.21 \pm 0.10$, calibrated using X-ray observations \citep{benson13}.  Given that the Phoenix cluster has the most-cuspy X-ray surface brightness profile of the SPT clusters with {\it Chandra} X-ray follow-up (McDonald et al. 2013), which includes $\sim$90 clusters, we expect the bias in $\zeta$ to represent an extreme example.  Therefore, the scatter in $\zeta$ given mass due to cool cores should be a factor of several below the overall scatter in $\zeta$ and its current measurement uncertainty.  The uncertainty in scatter has a negligible effect on the cosmological constraints from current SZ cluster surveys (e.g., \citet{benson13}), so the additional scatter from the effect of cool cores will be even less significant.
 
	Although we have focused on the SPT SZ survey, our results should be applicable to other SZ surveys, e.g. Planck, ACT, etc. Broadly speaking, we expect less bias in a survey with lower angular resolution and greater bias in a higher-resolution survey, since convolving a wide survey beam function with a cuspy SZ signal will smooth it to a less cuspy signal. 

\section{Conclusion}
	Using extensive Monte Carlo simulations of galaxy clusters and mock SPT observations, we have estimated the SZ bias due to cool cores in relaxed, massive systems. By doing so, we have constrained the cosmological and astrophysical bias due to the presence of cool cores and radio-loud AGN in SZ surveys. We find that the bias from cool cores is no larger than $\sim 10 \%$ for $z > 0.1$ systems, and for typical high redshift objects $(z \gtrsim 1, r_{500}\lesssim 800\, \mathrm{kpc}, \Delta \gamma \lesssim 1)$ the bias is at the percent level. Further, the presence of radio-loud sources in cool cores should reduce the overall bias, though at low redshifts $z\lesssim 0.3$, the bias from radio-loud point sources should dominate any cool core bias.

	Our results support the long-asserted claim that an SZ-selected sample of galaxy clusters is a robust cosmological probe: though we observe a small bias in the mass estimator, the magnitude of the bias is much smaller than the typical scatter in the mass relationship.

	We provide estimates of the SZ bias as a function of redshift, mass, and cuspiness, parameters that are model-independent, as well as the radio bias. One can estimate the bias of a given system easily by plugging in values to our function $\Delta\zeta/\zeta = f(z, r_{500}, \gamma)+g(L_{1.4},z,M_{500})$ where $f(z, r_{500}, \gamma)$ is given by equation (\ref{big}) and $g(L_{1.4},z,M_{500})$ is given by equation (\ref{eqn:agn}). By quantifying the cool core bias, astrophysical constraints on cool core properties from SZ surveys can now be more reliably interpreted. In particular, constraints on the cool core fraction from an SZ-selected dataset are only subject to a systematic bias of order one percent, a significant reduction over X-ray selection. Since there is a stringent upper limit on the redshift evolution of the cool core fraction bias, we can now confidently say that almost all observed evolution in the cool core fraction reflects genuine cool core evolution. With the arrival of results from Planck, SPT, and the Atacama Cosmology Telescope and others, SZ surveys might be ideal for studying the mysterious balance between heating and cooling.
\bibliographystyle{apj}
\bibliography{biblio}{}
\end{document}